\documentclass[pra,notitlepage,showpacs,superscriptaddress,amsfonts,amsmath,floatfix]{revtex4-1}
\usepackage{graphicx}
\usepackage{amsmath}
\usepackage{braket}
\usepackage{color}
\graphicspath{ {./images/} }
\begin{document}

\title{Experimental Demonstration of a Synthetic Lorentz Force\\
by Using Radiation Pressure}

\author{N. \v{S}anti\'{c}}
\affiliation{Department of Physics, University of Zagreb, Bijeni\v{c}ka c. 32, 10000 Zagreb, Croatia}
\affiliation{Institute of Physics, Bijeni\v{c}ka c. 46, 10000 Zagreb, Croatia}
\author{T.~Dub\v{c}ek}
\affiliation{Department of Physics, University of Zagreb, Bijeni\v{c}ka c. 32, 10000 Zagreb, Croatia}
\author{D.~Aumiler}
\affiliation{Institute of Physics, Bijeni\v{c}ka c. 46, 10000 Zagreb, Croatia}
\author{H.~Buljan}
\email{hbuljan@phy.hr}
\affiliation{Department of Physics, University of Zagreb, Bijeni\v{c}ka c. 32, 10000 Zagreb, Croatia}
\author{T.~Ban}
\email{ticijana@ifs.hr}
\affiliation{Institute of Physics, Bijeni\v{c}ka c. 46, 10000 Zagreb, Croatia}

\date{\today}

\begin{abstract}
Synthetic magnetism in cold atomic gases opened the doors to many exciting novel 
physical systems and phenomena. Ubiquitous are the methods used for the creation 
of synthetic magnetic fields. 
They include rapidly rotating Bose-Einstein condensates employing the analogy 
between the Coriolis and the Lorentz force, and laser-atom interactions 
employing the analogy between the Berry phase and the Aharonov-Bohm phase. 
Interestingly, radiation pressure - being one of the most common 
forces induced by light - has not yet been used for synthetic magnetism. 
We experimentally demonstrate a synthetic Lorentz force, based on the radiation 
pressure and the Doppler effect, by observing the centre-of-mass motion of a 
cold atomic cloud. 
The force is perpendicular to the velocity of the cold atomic cloud, and 
zero for the cloud at rest. 
Our novel concept is straightforward to implement in 
a large volume, for a broad range of velocities, 
and can be extended to different geometries. 
\end{abstract}

\maketitle
\let\newpage\relax

\section*{Introduction}

Experiments on synthetic magnetic/gauge fields for neutral atoms 
~\cite{Madison2000, Abo2001, Lin2009, LeB2012, Aidelsburger2011, Struck2012, 
Miyake2013, Aidelsburger2013, Aidelsburger2014, Jotzu2014, Ray2014}
have enabled realizations of the Hall effect~\cite{LeB2012}, famous Hamiltonians such 
as the Harper~\cite{Miyake2013, Aidelsburger2013} and the Haldane Hamiltonian~\cite{Jotzu2014}, 
intriguing topological effects~\cite{Aidelsburger2014,Jotzu2014}, and 
the observation of synthetic Dirac monopoles~\cite{Ray2014}. 
There are a few recent reviews on this promising theoretical and experimental progress 
in synthetic magnetic/gauge fields~\cite{Bloch2012, Dal2011, Goldman2014, Cooper2008}. 
The first implementation of synthetic magnetism was in rapidly rotating Bose-Einstein condensates (BECs), 
employing the analogy between the Lorentz force and the Coriolis force~\cite{Madison2000,Abo2001}.
The methods based on laser-atom interaction employ the analogy between 
the Berry phase in atomic systems~\cite{Dum1996}, and the Aharonov-Bohm phase for charged 
particles~\cite{Dal2011}. 
The first of them was realized in the NIST group 
with spatially dependent Raman optical coupling between internal 
hyperfine atomic states in bulk BECs~\cite{Lin2009}. 
Methods generating synthetic magnetic fields in optical lattices 
engineer the complex tunnelling matrix elements between lattice 
sites~\cite{Aidelsburger2011,Struck2012,Miyake2013,Aidelsburger2013,Aidelsburger2014}. 
They include shaking of the optical lattice~\cite{Struck2012}, 
laser assisted tunnelling in optical superlattices 
realizing staggered synthetic magnetic fields~\cite{Aidelsburger2011}, 
in tilted lattices realizing homogeneous fields~\cite{Miyake2013,Aidelsburger2013},
and an all-optical scheme which enables flux rectification in 
optical superlattices~\cite{Aidelsburger2014}. 
Interestingly, radiation pressure has not yet been used among the methods  
for synthetic magnetism, that is, to create the analogue of the Lorentz force.
Here we experimentally demonstrate the synthetic Lorentz force based on 
radiation pressure in cold atomic gases. 
We measure the dependence of the transverse radiation pressure force 
(analogous to the transverse Hall deflection) on the velocity 
of a cold atomic cloud by observing the centre-of-mass (CM) motion. 
The observed force is perpendicular to the velocity, 
and zero for the atomic cloud at rest. 
This concept based on radiation pressure, theoretically 
proposed in Ref.~\cite{Dubcek2014}, is straightforward to implement in 
a large volume (e.g., volumes 1~mm$^3$-1~cm$^3$ are easily accessible)~\cite{Met1999}, 
is applicable for a broad range of velocities, and can be extended 
to different geometries. 
The main reason for the absence of radiation pressure from the previously used methods 
of synthetic magnetism is the associated heating due to spontaneous emission. 
However, this is not an obstacle for atomic gases cooled and trapped 
in a Magneto-Optical Trap (MOT), where our experiments are performed. 

\section*{Results}


\begin{figure}[h]
\centerline{
\mbox{\includegraphics[width=0.95\textwidth]{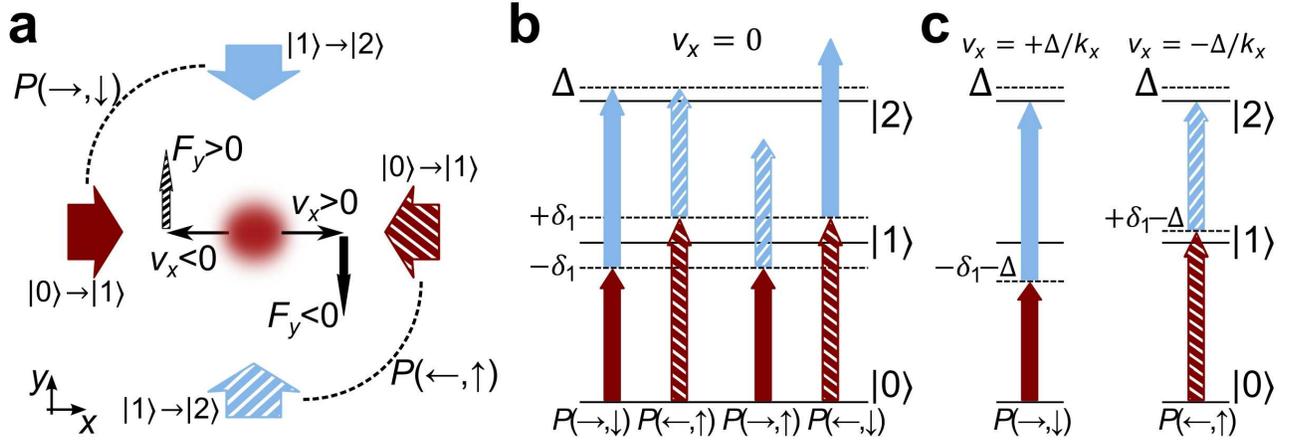}}
}
\caption{The scheme used to obtain the synthetic Lorentz force via radiation pressure. 
It is based on two-step two-photon transitions involving three atomic levels. 
(a) Two counter-propagating laser beams on the $x$-axis (red arrows)
drive the $\ket{0}\rightarrow \ket{1}$ transition, while the   
$\ket{1}\rightarrow \ket{2}$ transition is driven by 
two counterpropagating beams on the $y$-axis (blue arrows). 
The transverse radiation pressure force arising from the $\ket{1}\rightarrow \ket{2}$ 
transition $F_y$, depends on the velocity $v_x$ as indicated. 
(b) Four possible excitation pathways and a sketch of the 
detuning values: 
$P(\rightarrow,\downarrow)$ denotes absorption of a $\ket{0}\rightarrow \ket{1}$ 
photon going towards the positive $x$-direction, followed by absorption 
of a $\ket{1}\rightarrow \ket{2}$ photon travelling towards the negative $y$-direction, and so on. 
The total detuning value for $P(\rightarrow,\downarrow)$ and $P(\leftarrow,\uparrow)$ is $\Delta$, 
whereas it is much larger in magnitude for $P(\rightarrow,\uparrow)$ and $P(\leftarrow,\downarrow)$;
the two latter pathways are thus negligible in this configuration.  
(c) The Doppler shift provides $F_y$ as a function of $v_x$ as sketched in (a): 
pathway $P(\rightarrow,\downarrow)$ for an atom with positive 
velocity $v_x=+\Delta/k_x$ ($v_y=0$) is on resonance, providing negative $F_y$. 
Likewise, $P(\leftarrow,\uparrow)$ is on resonance for an atom with negative velocity 
$v_x=-\Delta/k_x$, providing positive $F_y$. 
See text for details. 
}
\label{sketch}
\end{figure}


{\bf The basic idea.} 
The idea behind our experiment is to drive two-step two-photon 
transitions between three atomic levels, 
$\ket{0}\rightarrow \ket{1}\rightarrow \ket{2}$, with mutually perpendicular 
laser beams as illustrated in Fig. \ref{sketch}.
Two counterpropagating laser beams aligned with the $x$-axis drive the 
$\ket{0}\rightarrow \ket{1}$ transition, whereas the $\ket{1}\rightarrow \ket{2}$ 
transition is driven by counterpropagating beams aligned on the $y$-axis. 
Due to the Doppler effect and the perpendicular configuration of the laser beams, 
both components of the radiation pressure force depend on both 
components of the atomic velocity: $F_x=F_x(v_x,v_y)$ and $F_y=F_y(v_x,v_y)$. 
This gives us the opportunity to design the detuning values of our 
lasers such that $F_y$ is positive/negative for atoms with 
negative/positive velocity component $v_x$, and that 
the total force is zero for an atom at rest: ${\bf F}({\bf v}={\bf 0})={\bf 0}$. 
These are the characteristics of the synthetic Lorentz force that we experimentally demonstrate.

The design of the detuning values of the 
lasers is crucial in obtaining the desired result. 
The beams driving the first step of the transition $\ket{0} \rightarrow \ket{1}$ 
are detuned by the same magnitude, but with the opposite sign.
The one towards the positive $x$-direction is red-detuned by 
$\delta_{\rightarrow}=-\delta_1<0$, while the other is blue-detuned by $\delta_{\leftarrow}=\delta_1>0$. 
Their intensities are equal. Thus, if just these two lasers were present, 
the net force on atoms (of any velocity) would be zero.  
However, the population of level $\ket{1}$ would depend on the 
velocity $v_x$, which implies that the rate of transitions $\ket{1} \rightarrow \ket{2}$
giving the transverse force will depend on $v_x$. 
The detuning values of the beams driving the second step of the transition, 
$\ket{1} \rightarrow \ket{2}$ are denoted by $\delta_{\uparrow}$ and $\delta_{\downarrow}$, for 
the beam propagating in the positive and negative $y$ direction, respectively. 
For now, let us set these values such that $\delta_{\rightarrow}+\delta_{\downarrow}=
\delta_{\leftarrow}+\delta_{\uparrow}=\Delta>0$, 
as indicated in Fig.~\ref{sketch}(b).

The two-step two-photon transitions, where absorption of a 
$\ket{1} \rightarrow \ket{2}$ photon follows absorption of a 
$\ket{0} \rightarrow \ket{1}$ photon with perpendicular momentum, 
yield the synthetic Lorentz force via momentum transfer from photons to atoms. 
Given the fact that we have two counterpropagating beams for each transition, 
we have four excitation pathways for the two-step two-photon 
transition, denoted by $P(\rightarrow,\uparrow)$, $P(\leftarrow,\uparrow)$, 
$P(\rightarrow,\downarrow)$, and $P(\leftarrow,\downarrow)$, see Fig. \ref{sketch}. 
The arrows correspond to the direction of the photon's momentum, for example, 
$P(\rightarrow,\downarrow)$ denotes the pathway where absorption of a 
photon travelling in the $+x$ direction is followed by absorption 
of a photon in the $-y$ direction and so on. 
Since the detuning magnitudes of the first step are identical for 
all pathways ($|\delta_\rightarrow|=\delta_\leftarrow$), the relevant 
quantity is the total detuning for the two-step two-photon transitions. 
It is important to note that the total detuning $\Delta$ for pathways 
$P(\rightarrow,\downarrow)$ and $P(\leftarrow,\uparrow)$, is much smaller 
in magnitude than the detuning values of $P(\rightarrow,\uparrow)$ and 
$P(\leftarrow,\downarrow)$. 
The last two are thus negligible in the setup of Fig.~\ref{sketch}.

To understand the origin of the synthetic Lorentz force, we take into account the Doppler shift. 
For an atom moving along the $x$-axis with velocity $v_x$ $(v_y=0)$, the 
$P(\leftarrow,\uparrow)$ pathway is detuned by $\Delta+k_x v_x$, i.e., it is 
on resonance when $v_x=-\Delta/k_x<0$. Because photons from the 
second step of $P(\leftarrow,\uparrow)$ impart momentum towards the positive $y$-direction, 
there will be a positive force $F_y$ for atoms with negative $v_x$. 
In the same fashion, the $P(\rightarrow,\downarrow)$ pathway will be on resonance when 
$v_x=+\Delta/k_x>0$ yielding negative $F_y$ for positive $v_x$.

{\bf The experiment.}
In the experiment, we use cold $^{87}$Rb atoms. 
For the $\ket{0}\rightarrow \ket{1}$ transition we use the D2 transition in $^{87}$Rb:
$|5S_{1/2};F=2\rangle \rightarrow |5P_{3/2};F'=3\rangle$ at 780~nm~\cite{Steck}. 
The easily accessible transition, $|5P_{3/2};F'=3\rangle \rightarrow |5D_{5/2};F''=4\rangle$ 
at 776~nm~\cite{Sheng2008}, is used for the second step $\ket{1}\rightarrow \ket{2}$. 
The linewidths of the states are $2\pi\times 6.1$~MHz for the $|5P_{3/2};F'=3\rangle$ state~\cite{Steck}, 
and $2\pi\times 0.66$~MHz for the $|5D_{5/2};F''=4\rangle$ state~\cite{Sheng2008}.

The $^{87}$Rb atoms are cooled and trapped in a standard glass vapour cell magneto-optical trap (MOT), 
arranged in a retro-reflected configuration. In this configuration, three orthogonal 
retro-reflected beams are used to create the total of six beams needed for 
the MOT. A pair of anti-Helmholtz coils provides a quadrupole magnetic field, 
which together with the laser beams creates a trapping potential  
(for example, see Ref.~\cite{Met1999}).  
Fluorescence imaging of the cloud is performed with a camera aligned along the $z$-axis. 
In typical experimental conditions we obtain a cloud of $0.4$~mm in diameter, which 
contains about $10^{8}$ atoms of $^{87}$Rb, at a temperature of $50$~$\mu$K
(for details of the experimental setup see Methods).
The four beams implementing the synthetic Lorentz force, arranged as in 
Fig.~\ref{sketch}, are of much smaller intensity than the MOT beams. 
Therefore, they are negligible when the MOT beams are ON. For the 
experimental detection of the synthetic Lorentz force, we 
turn the MOT beams OFF, as described in detail below.

We need to measure the transverse force $F_y$ in dependence of the 
velocity $v_x$ of the atomic cloud. Thus, we must prepare 
a cloud with a given centre of mass (CM) velocity. 
For this purpose we use an additional pair of current coils that produce a bias 
magnetic field along the symmetry axis of the anti-Helmholtz coils, $x$ in our notation here.
The bias field moves the centre of the trap (the point where ${\bf B}=0$), which 
displaces the cloud approximately $1$~mm along the $x$-axis.

The measurement protocol is as follows. 
(i) We load the trap with the bias magnetic field on. 
(ii) At $t=-2$~ms we reverse the bias magnetic field by reversing the current, 
which suddenly shifts the centre of the trap. This introduces a force 
on the cloud due to the MOT beams. 
During the next $2$~ms the cloud accelerates along the $x$-axis towards the 
new trap centre. 
(iii) Next we turn off the MOT cooling laser and all real magnetic fields. 
This moment corresponds to $t=0$ in our presentation. 
The system is now simplified because Zeeman splitting of 
hyperfine levels is absent and the radiation force left, arising solely 
from the lasers implementing the synthetic Lorentz force, is not spatially dependent 
(it is only velocity-dependent). 
(iv) The cloud starts to expand because 
the trapping is absent, but it also moves in the $xy$ plane due to both the 
initial velocity, $v_x(t=0)$, and the radiation pressure force. 
(v) After some delay time $t$, the cooling laser is suddenly turned on and the 
cloud is imaged with the camera. From the trajectory traversed by the CM 
of the cloud $(x(t),y(t))$, we can find the force acting on the atoms. 
If we wish to image a cloud initially at rest we skip steps 
(i) and (ii). 
For a given delay time $t$, we repeat the measurement protocol 
20 times in identical conditions, and subsequently average to 
obtain $x(t)$ and $y(t)$. The gravity is along the $z$-axis in our system; 
free fall of atoms due to gravity does not affect the motion in the $xy$ plane. 
It should be stated that we perform these measurements first with 
$\ket{1}\rightarrow \ket{2}$ lasers OFF, and then with these lasers 
ON, keeping all other parameters identical. The difference in the path $y(t)$, 
with lasers $\ket{1}\rightarrow \ket{2}$ ON and OFF, gives us the 
transverse motion due to the synthetic Lorentz force.

 
 \begin{figure}[h]
 \centerline{
 \mbox{\includegraphics[width=0.375\textwidth]{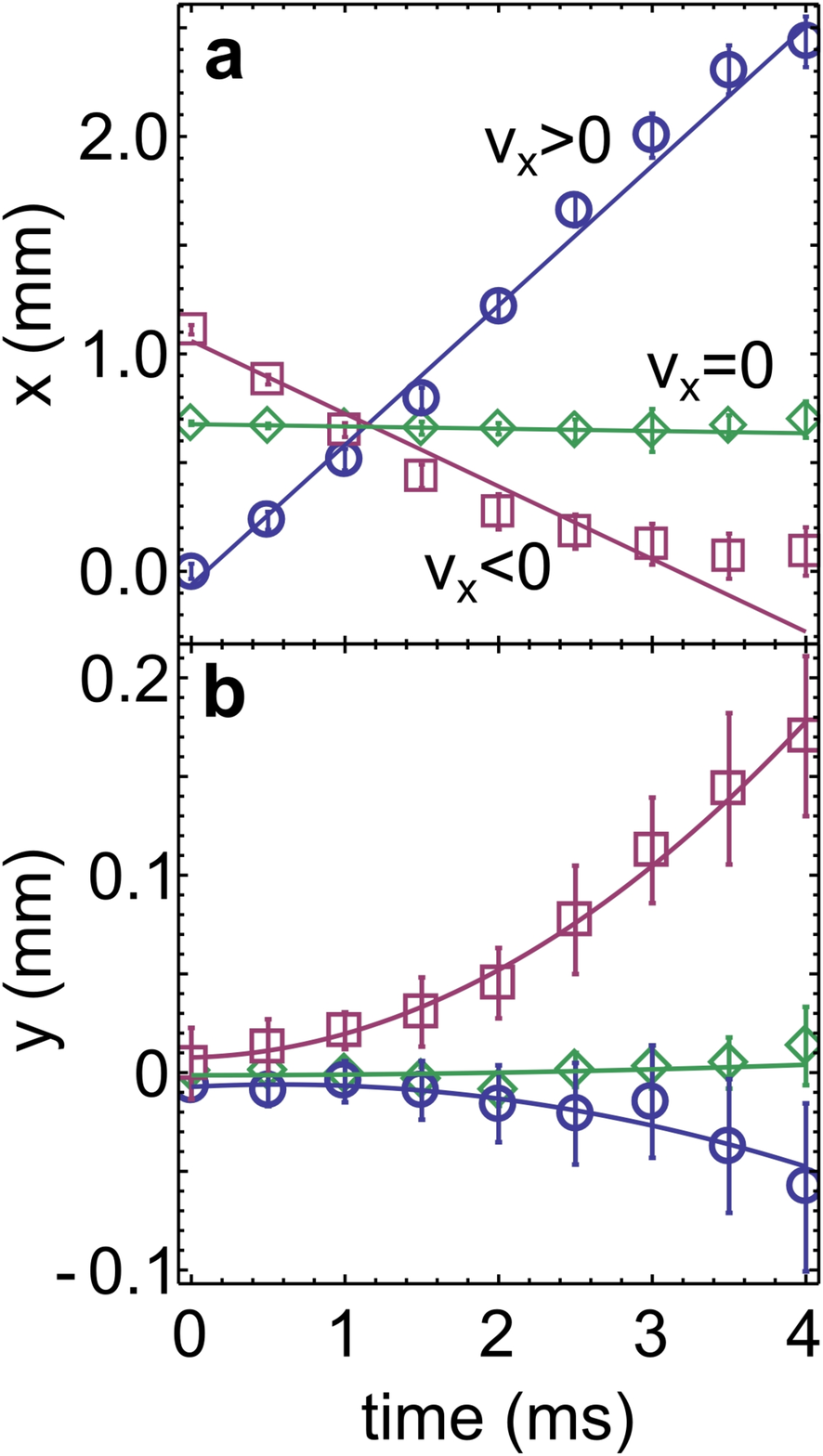}}
 }
 \caption{The trajectories of the CM of the atomic cloud in the presence of the synthetic Lorentz force. 
 (a) $x(t)$, and (b) $y(t)$ for three different initial velocities, 
 $v_x=0.6$~m/s $>0$ (circles), $v_x=-0.3$~m/s $<0$ (squares), 
 and $v_x=0$~m/s (diamonds); initial component of $v_y=0$ in all measurements.
 Accelerating motion along $y$ is the signature of the transverse force $F_y$,
 which depends on $v_x$. The lines are fitted to the experimental data, 
 linear fits for $v_x(t)$, and quadratic for $v_y(t)$.  
 }
 \label{results}
 \end{figure}
 

{\bf The experimental proof of the synthetic Lorentz force by radiation pressure. }
The results of the experiment are illustrated in Fig.~\ref{results}. 
We show the trajectory of the cloud $(x(t),y(t))$, in the presence of the 
synthetic Lorentz force, for three initial velocities: 
$v_x=-0.3$~m/s (squares), $v_x=0$~m/s (diamonds), and $v_x=0.6$~m/s (circles);
$v_y=0$ at $t=0$ in each run of the experiment. 
There is a difference in the magnitude of the initial $v_x$ for the positive 
and negative velocity [circles and squares in Fig.~\ref{results}(a)], which is 
a result of our MOT retro-reflected geometry, and the way we accelerate the 
cloud in step (ii) of the protocol. In order to prepare a cloud with positive 
$v_x$, the cloud is accelerated with the incoming MOT beam (coming from the laser 
side of the setup), whereas acceleration in the opposite direction is performed 
with the reflected beam which has smaller intensity. The reflected beam intensity 
is smaller due to the losses, which are a result of the passage of the incoming 
beam through the dense cloud (absorption), and partially due to reflection.
Consequently, the negative initial velocity is smaller than the positive velocity.

For the lasers implementing the synthetic Lorentz force, we use the following detuning values:
for the first step at 780 nm $\delta_{\leftarrow}=-\delta_{\rightarrow}=2\pi\times 6$~MHz, 
and for the second step at 776 nm $\delta_{\uparrow}=-2\pi\times 3.5$~MHz 
and $\delta_{\downarrow}=2\pi\times 7.1$~MHz. 
The two operational two-step pathways are $P(\leftarrow,\uparrow)$ and 
$P(\rightarrow,\downarrow)$, whereas the other two are far from resonance. 
The intensities of the beams used
are $I_{780}=0.060$~mW/cm\textsuperscript{2} 
and $I_{776}=2.9$~mW/cm\textsuperscript{2}, giving Rabi frequencies 
$\Omega_{780}=2\pi\times 1.2$~MHz and $\Omega_{776}=2\pi\times 0.94$~MHz.

By inspection of Fig.~\ref{results}, we see that the cloud travels along the $x$-axis by inertia, 
whereas it accelerates along the $y$-axis due to the synthetic Lorentz force. 
The direction of the force depends on the sign of the velocity $v_x$ 
(negative/positive $v_x$ gives positive/negative $F_y$), and the force is zero for a cloud at rest. 
We observe an asymmetry in the force $F_y$, for the positive and negative velocity. 
In order to justify the exact choice of the detuning values $\delta_{\uparrow}$ 
and $\delta_{\downarrow}$, and further investigate 
the observed asimmetry in the measured synthetic Lorentz force, 
we perform measurements in a slightly simplified configuration.


\begin{figure}[h]
\centerline{
\mbox{\includegraphics[width=0.375\textwidth]{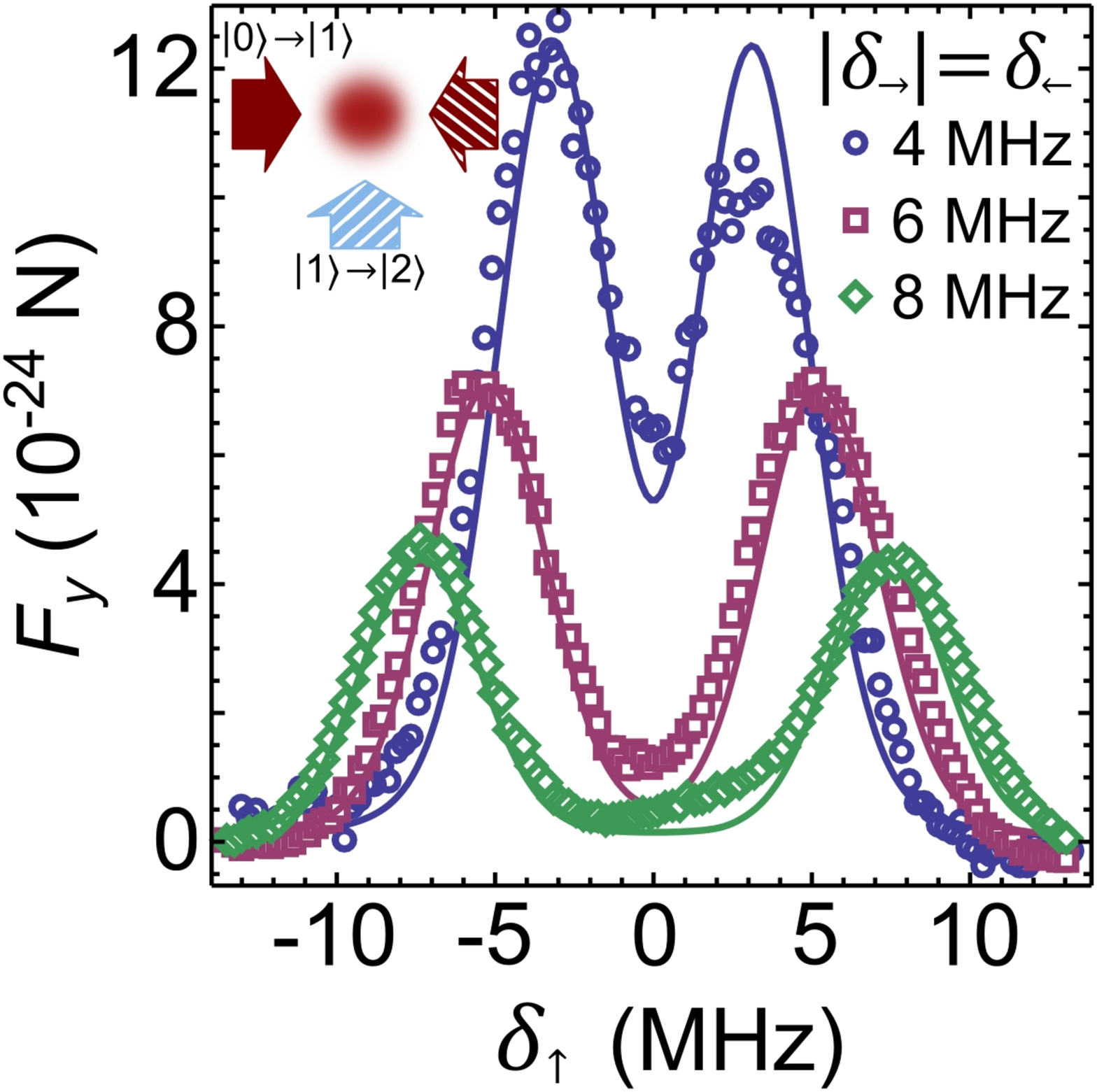}}
}
\caption{Frequency scan of the two-step two-photon resonance
in the auxiliary configuration. 
The calculated (solid lines) and measured force $F_y$ as a function of  
$\delta_{\uparrow}$ for the excitation with just three beams as shown. 
Measurements are performed for the cloud with initial velocity zero. 
The plots show resonances for three values of the detuning 
$\delta_{\leftarrow}=-\delta_{\rightarrow}=2\pi \times 4$~MHz (circles), 
$6$~MHz (squares) and $8$~MHz (diamonds). 
}
\label{scan}
\end{figure}


{\bf Measurements in an auxiliary configuration. }
We inspect the force along the $y$ direction arising from the two-step two-photon 
resonances, by using a configuration with three laser beams illustrated in Fig. \ref{scan}.
We block the beam pointing towards the negative $y$-direction,
and measure $F_y$ arising from the remaining beam (the positive $y$-direction), 
which drives the $\ket{1} \rightarrow \ket{2}$ transition. 
The force is measured for an atomic cloud with velocity zero, as a function of the 
detuning $\delta_{\uparrow}$, see Fig. \ref{scan}. 
Measurements are done for three different detuning values of the first-step 780~nm beams 
$\delta_{\leftarrow}=-\delta_{\rightarrow}=2\pi\times 4, 6, 8$~MHz. 
The intensities of the lasers driving the transitions are 
now $I_{780}=0.043$~mW/cm\textsuperscript{2} 
and $I_{776}=4.8$~mW/cm\textsuperscript{2}, giving Rabi frequencies 
$\Omega_{780}=2\pi\times 1.1$~MHz and $\Omega_{776}=2\pi\times 1.2$~MHz.
The two maxima in Fig. \ref{scan} are profiles of the two-step two-photon resonances: 
the peak in the vicinity of $\delta_{\uparrow}=-\delta_{\rightarrow}>0$ corresponds to the 
excitation pathway $P(\rightarrow,\uparrow)$, and the peak close to 
$\delta_{\uparrow}=-\delta_{\leftarrow}<0$ corresponds to $P(\leftarrow,\uparrow)$.

Solid lines in Fig. \ref{scan} show the theoretically calculated 
profiles of $F_y$ (see Methods for details of the theoretical calculation). 
The agreement between theory and experiment is evident. 
All parameter values used in the theoretical calculation are taken from the experiment, 
except that the Rabi frequencies are reduced by $20$\%. 
This is reasonable because in the experiment, the absorption of laser beams across 
a dense atomic cloud reduces their intensity~\cite{Kaiser}.

It should be pointed out that the peaks in $F_y(\delta_{\uparrow})$ are slightly displaced 
from the values $\delta_{\uparrow}=\pm|\delta_{\rightarrow}|$, towards $\delta_{\uparrow}=0$. 
For example, for $|\delta_{\rightarrow}|$ at $6$~MHz, the maxima are at $\pm 5.3$~MHz. 
Moreover, the FWHM of the peaks is larger than the linewidth of the state $\ket{2}$. 
These two observations are a consequence of the laser linewidth, which is larger than 
the linewidth of the state $\ket{2}=|5D_{5/2};F''=4\rangle$ (see Methods).
The finite linewidth of $\ket{0}\rightarrow \ket{1}$ lasers at 780~nm distorts the 
peaks as follows: the side of the peak closer to 
$\delta_{\uparrow}=0$ is lifted up in comparison the opposite side of the peak. 
This follows from the fact that the two-step two-photon resonance is 
stronger when $|\delta_{\rightarrow}|$ is closer to zero, 
which is evident from Fig. \ref{scan}. 
The exact positions of the peaks at $\pm 5.3$~MHz, explain the chosen detuning 
values for the second step at 776~nm, which were used to obtain Fig.~\ref{results}: 
$\delta_{\uparrow}=-2\pi\times 3.5$~MHz and $\delta_{\downarrow}=2\pi\times 7.1$~MHz. 
They are chosen such that the operational pathways are effectively equally detuned
from the two-step two-photon resonance:
$P(\rightarrow,\downarrow)$ is detuned by $2\pi\times (-5.3+7.1)$~MHZ$=2\pi\times 1.8$~MHZ, and 
$P(\leftarrow,\uparrow)$ is detuned by $2\pi\times (+5.3-3.5)$~MHZ$=2\pi\times 1.8$~MHZ.


\begin{figure}[h]
\centerline{
\mbox{\includegraphics[width=0.375\textwidth]{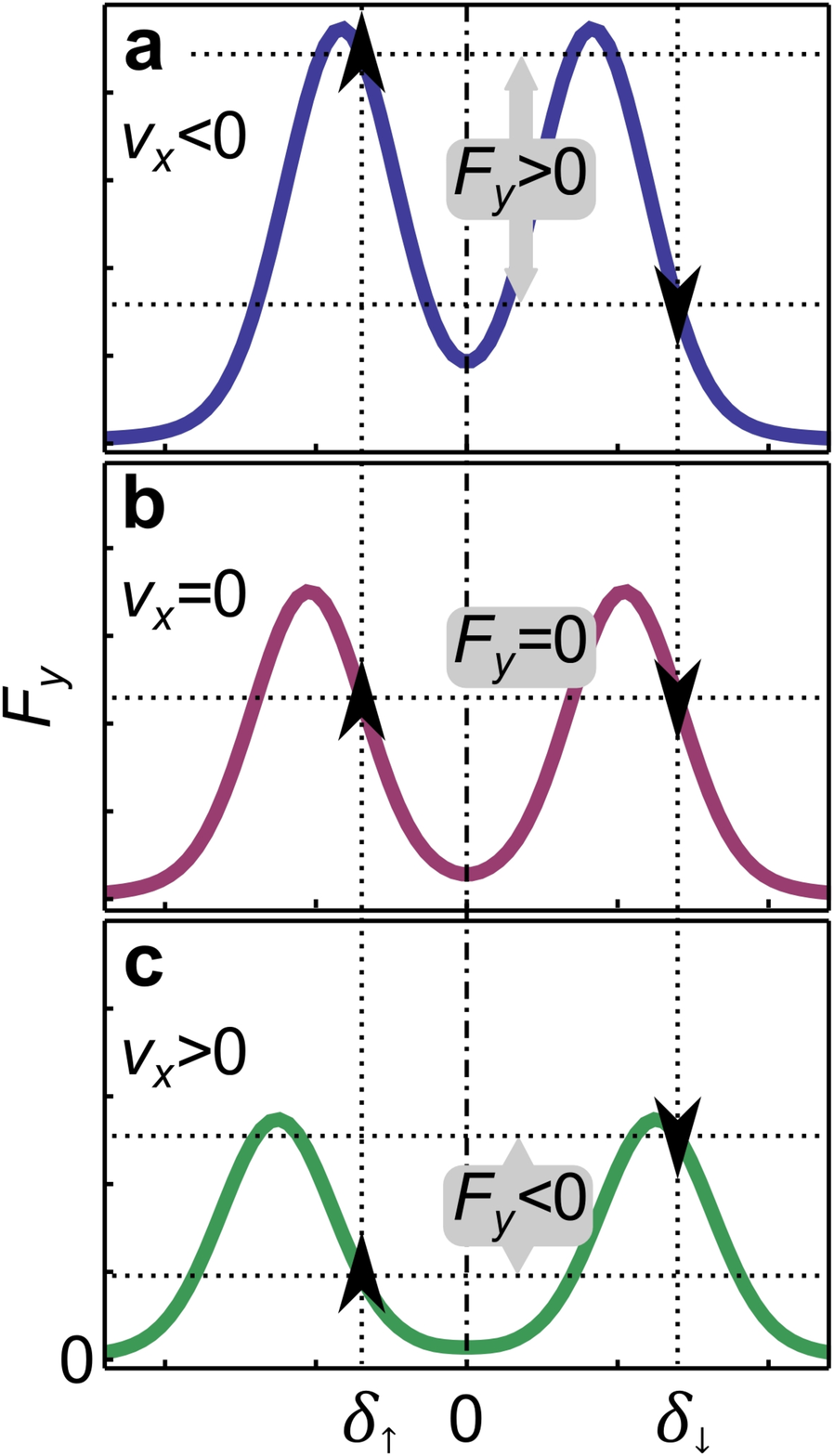}}
}
\caption{Interpretation of the synthetic Lorentz force 
from Figs. \ref{sketch} and \ref{results}, via two-step two-photon 
resonances presented in Fig. \ref{scan}. 
Sketch of the resonant peaks that would be obtained in the setup shown 
in Fig. \ref{scan} for three atomic velocities: (a) $v_x<0$, (b) $v_x=0$, and 
(c) $v_x>0$. 
Vertical dotted lines illustrate the values of the detuning used 
for the lasers aligned on the $y$-axis in Fig. \ref{sketch}.
Arrows denote the direction of the force exerted by those lasers, 
and illustrate the way $F_y$ observed in Fig. \ref{results} 
arises as a function of velocity $v_x$. 
See text for details. 
}
\label{concept}
\end{figure}


\section*{Discussion}

Suppose that we repeat measurements corresponding to Fig.~\ref{scan}, 
but for an atomic cloud with mean velocity $v_x$ different from zero. 
The results of such measurements would be identical as for $v_x=0$, but 
the positions of the peaks would correspond to 
the Doppler shifted detuning values $\delta_{\rightarrow}-kv_x$ 
and $\delta_{\leftarrow}+kv_x$. Thus, because detuning can be mapped to 
velocity space, Fig.~\ref{scan} can be reinterpreted as measurements 
for a fixed value of $\delta_{\leftarrow}=-\delta_{\rightarrow}$, and for 
three different velocities $v_x<0$ (4~MHz), $v_x=0$ (6~MHz), and $v_x>0$ (8~MHz). 
This is sketched in Figure ~\ref{concept}, where we see that 
the two peaks separate (approach) each other for $v_x>0$ 
($v_x<0$, respectively).

We use Fig.~\ref{concept} for a detailed explanation of the synthetic Lorentz force 
measured in Fig.~\ref{results}. 
In measurements shown in Fig.~\ref{results}, we have $\delta_{\uparrow}<0$, 
which means that the positive force $F_y>0$ in Fig.~\ref{results}
results from the left resonance peak in Fig.~\ref{concept}. 
Likewise, because $\delta_{\downarrow}>0$ was used for Fig.~\ref{results}, 
the negative force $F_y<0$ results from the right resonance peak in Fig.~\ref{concept}. 
The transverse force $F_y$ measured in Fig.~\ref{results}, 
can be approximately identified with $[F_y (\delta_{\uparrow}) - F_y (\delta_{\downarrow})]_{v_x}$, 
as illustrated in Fig.~\ref{concept}.

The choice of $\delta_{\uparrow}$ and $\delta_{\downarrow}$ 
in Fig.~\ref{results} is such that the strength of the forces arising from the two 
peaks balance each other, giving $F_y=0$ for $v_x=0$ [Fig.~\ref{concept}(b)]. 
Moreover, $\delta_{\uparrow}$ and $\delta_{\downarrow}$ are on the slopes of the 
two peaks in Fig. \ref{concept}(b) for an atom with $v_x=0$,
where the maximal magnitude of $\Delta F_y/\Delta v_x$ is expected 
(detuning translates into velocity space via Doppler effect). 
For an atom with $v_x<0$ [see Fig.~\ref{concept}(a)], the two peaks approach 
each other, yielding greater force from the left peak, 
which results in $F_y>0$ for $v_x<0$, and the opposite for 
$v_x>0$ which yields $F_y<0$ as shown in Fig.~\ref{concept}(c).

It should be pointed out that our scheme is inherently asymmetric. 
The intensity of the resonance peaks shown in Fig.~\ref{concept} 
decreases when they separate (for $v_x>0$), in contrast to when they approach each other 
(for $v_x<0$). Thus, the net force along $y$ is larger for negative $v_x$, 
than for the velocity of the same magnitude but with a positive sign. 
By using a scheme with four atomic levels~\cite{Dubcek2014}, one could remedy 
the asymmetry in the force, present in the three-level scheme. 
In addition, the initial density of the atomic cloud is not identical 
prior to acceleration in the positive/negative direction due to 
the slight asymmetry of the displaced trapping potential.  
This also affects the transverse force due to absorption and multiple 
scattering~\cite{Kaiser}. 
The inherent asymmetry in the force $F_y(v_x)$ and in the initial 
density of the cloud, is reflected in the asymmetry of the 
observed motion in Fig.~\ref{results} for $v_x<0$ and $v_x>0$.
Importantly, all of these details only quantitatively affect the results, 
but not qualitatively.

\section*{Conclusion}

The presented experiment, which demonstrates the synthetic Lorentz force
by using radiation pressure, is performed in a classical cold 
atomic gas, prepared in a MOT, where the heating due to spontaneous emission 
does not prevent the observations. 
The concept of a force is natural in our classical laser cooled system~\cite{Met1999}. 
We would like to emphasize that our method is entirely different from the Berry phase approach, 
where the connection with the Lorentz force can be made in a semiclassical 
approximation~\cite{Cheneau2008}.

In the outlook, we foresee many intriguing novel experiments based on the presented method. 
First, one could develop experiments using more sophisticated schemes, involving 
more atomic levels (see Ref.~\cite{Dubcek2014}), to create a uniform synthetic 
magnetic field. Next, we plan to adjust our system for the observation of 
the predicted rotation of the cloud during expansion~\cite{Dubcek2014}. 
One of the goals of this research is to build up a toroidal trap for cold atoms 
with a toroidal synthetic magnetic field, which holds potential to emulate the plasma 
in a tokamak. The proposed concept could be used for velocity selection of atomic beams, 
or for developing a novel kind of mass spectrometer for neutral atoms. 
We believe that our concept or an analogous scheme could be applicable 
in other systems, such as suspended nanoparticles.

\section*{Methods}

{\bf Experimental setup.}
The $^{87}$Rb MOT is set up in the standard $\sigma^+$ $\sigma^-$ retro-reflected configuration, 
with beam diameters of 2 cm. The trap is vapour loaded in a glass cell which 
facilitates fast switching of the magnetic field. 
Cooling and repumper lasers are external cavity diode lasers (ECDL) 
delivering total powers of 80 mW and 20 mW, respectively. 
The cooling laser is typically 2$\pi\times$24~MHz (4$\Gamma$) red-detuned from the 
$|5S_{1/2};F=2\rangle \rightarrow |5P_{3/2};F'=3\rangle$ hyperfine transition. 
The repumper laser is in resonance with the $|5S_{1/2};F=1\rangle \rightarrow |5P_{1/2};F'=2\rangle$ 
hyperfine transition, thus keeping most of the population in the 
$|5S_{1/2};F=2\rangle$ ground level. The quadrupole magnetic field 
gradient is 13~G/cm. 
The number of atoms in the trap is deduced by measuring the cloud fluorescence with a calibrated photodiode.

For the implementation of the synthetic Lorentz force two additional 
ECDL lasers were introduced in the experiment: 
one at 780 nm driving $|5S_{1/2};F=2\rangle \rightarrow |5P_{3/2};F'=3\rangle$ transition, 
and other at 776 nm driving $|5P_{3/2};F'=3\rangle \rightarrow |5D_{5/2};F''=4\rangle$ transition. 
Each laser beam is split into two beams which are sent on the rubidium cloud 
in counter-propagating configurations as shown in Fig.~\ref{sketch}. 
Frequency and intensity control is done separately for each of the four laser beams with acoustic-optical modulators (AOM).
After being frequency shifted, we couple the beams to single mode polarization 
maintaining fibres, which ensures linear polarization (in the $z$-direction) 
and uniform intensity. 

All lasers used in the experiment are frequency stabilized by using modulation transfer 
spectroscopy.
We modulate the laser diode current to modulate the laser frequency, 
which effectively increases the laser linewidth. 
This additional broadening is taken into account in theoretical calculations. 
The linewidths were checked by heterodyne beating of two stabilized lasers 
with similar locking parameters.
For locking the laser tuned at 776~nm, we counter-propagate picked off beams from the 776~nm 
laser and the 780~nm laser through a heated $^{87}$Rb glass cell, where the 780~nm 
laser populates the $|5P_{3/2};F'=3\rangle$ level. We observe an absorption signal 
resulting from the $|5P_{3/2};F'=3\rangle \rightarrow |5D_{5/2};F''=4\rangle$ transition. 
This signal is mixed with the modulation signal from the 780~nm laser to obtain a frequency locking 
error signal. 
Therefore, there is no need to modulate the 776~nm laser frequency to stabilize it.

{\bf Theoretical calculation of the radiation pressure force.}
To calculate the force plotted in Fig. \ref{scan}, 
we first solve the optical Bloch equations (OBEs) to find the density matrix 
$\hat \rho$~\cite{Met1999}:
\begin{displaymath}
{\frac {d\rho_{nm}}{dt}}={-\frac{i}{\hbar}} 
\left[\hat H,\hat\rho \right]_{nm}-\gamma_{nm} \rho_{nm}, \qquad n\neq m,
\end{displaymath}
\begin{equation}
{\frac {d\rho_{nn}}{dt}}={-\frac{i}{\hbar}} 
\left[\hat H,\hat\rho \right]_{nn}+\sum_{E_m>E_n}{\Gamma_{nm} \rho_{mm}}-\sum_{E_m<E_n}{\Gamma_{mn} \rho_{nn}},
\end{equation}
where $n,m=0,1,2$.
The Hamiltonian ${\hat H}={\hat H}^{(0)}+{\hat H}^{(1)}$ describes a three-level system interacting 
with laser fields. ${\hat H}^{(0)}$ is represented by a diagonal matrix with elements 
$H^{(0)}_{00}=0$, $H^{(0)}_{11}=\hbar\omega_{01}$, and $H^{(0)}_{22}=\hbar(\omega_{01}+\omega_{12})$, 
whereas the interaction Hamiltonian is, in the dipole and the rotating wave 
approximation~\cite{Met1999},
\begin{align}
{\hat H}_{12}^* & = {\hat H}_{21} = 
-\frac{\hbar\Omega_{12}}{2} 
(
 e^{i (  k_xx - \omega_{01} t - \delta_\rightarrow t )}
+e^{i ( -k_xx - \omega_{01} t - \delta_\leftarrow  t )}
),
\nonumber \\
{\hat H}_{23}^* & = {\hat H}_{32} = 
-\frac{\hbar\Omega_{23}}{2} 
e^{i ( k_yy - \omega_{12}t - \delta_\uparrow t )} , 
\end{align}
and zero otherwise.
The linewidths, $\Gamma_{12}=2\pi\times 6.1$~MHz, $\Gamma_{23}=2\pi\times 0.66$~MHz, 
and the transition frequencies, $\omega_{01}=k_x c$ and $\omega_{12}=k_y c$, 
where $k_x=2\pi/780$~nm$^{-1}$ and $k_y=2\pi/776$~nm$^{-1}$, 
correspond to the experiment; $\Gamma_{13}=0$; 
the coherences are $\gamma_{ij}=\frac{1}{2}\left(\Gamma_{ii}+\Gamma_{jj}\right)$, 
where $\Gamma_{ii}=\sum_{j<i}{\Gamma_{ji}}$.
Because transition $\ket{0} \rightarrow \ket{1}$ is driven with two lasers of 
different detuning values $\delta_{\leftarrow}=-\delta_{\rightarrow}$, 
there is no stationary solution, and the level populations oscillate with the 
frequency $|\delta_{\rightarrow}|\sim 6$~MHz (the phase of these 
oscillations depends on $x$, but this is not reflected in the force). 
The force on an atom is calculated by employing the Ehrenfest theorem \cite{Met1999}, 
and averaging over the oscillations: 
$F_y'(\delta_\rightarrow,\delta_\leftarrow,\delta_\uparrow) = 
\langle - {\mbox Tr} [ {\hat \rho} \partial {\hat H}/ \partial y ] \rangle_t$.
The average over time $\langle \cdots \rangle_t$ is justified because the 
oscillations at frequency $|\delta_{\rightarrow}|\sim 6$~MHz are much faster than the characteristic scale 
for measuring the force, $\tau_F$, which is on the order of a few milliseconds.

The outlined procedure for calculating the force assumes that the lasers 
are perfectly monochromatic. However, there is finite laser linewidth 
that should be taken into account to quantitatively describe 
the two-step two-photon resonances in Fig. \ref{scan}, i.e., 
$F_y(\delta_\uparrow)$. 
The spectral profile of the diode laser can be described by a 
Gaussian in the frequency domain, 
$\propto \exp((\omega-\omega_0)^2/2\sigma^2)$.
In the experiment, the modulation transfer spectroscopy for laser locking is 
used, and the central laser frequency is modulated with the amplitude 
$d\approx 2\pi\times 1$~MHz and frequency $\eta=2\pi\times 14$~kHz:
$\tilde \omega_0 (t) = \omega_0 + d \sin \eta t$.
Thus, the laser spectral profile, over time-scales larger than $1/\eta$, is given by 
\begin{align}
p( \omega ; \omega_0)
&=\frac{p_0 \eta d}{\sigma\sqrt{2\pi^3}}
\int_{\omega_0-d}^{\omega_0+d}
\frac{\exp(-(\omega-\tilde \omega_0)^2 / 2\sigma^2)}
{\lvert\frac{d \tilde \omega_0}{dt}\rvert_{t(\tilde \omega_0)}}
d \tilde \omega_0
\nonumber \\
&= \frac{p_0}{\sigma\sqrt{2\pi^3}}
\int_{\omega_0-d}^{\omega_0+d}
\frac{\exp(-(\omega-\tilde \omega_0)^2 / 2\sigma^2)}
{\sqrt{1-\left(  \frac{\tilde\omega_0-\omega_0}{d} \right)^2}}
d \tilde \omega_0, 
\end{align}
where $p_0$ is the normalization factor.

The detuning values of $780$~nm lasers oscillate as follows: 
$\tilde\delta_{\rightarrow} = \delta_{\rightarrow} + d \sin \eta t$ and 
$\tilde\delta_{\leftarrow} = \delta_{\leftarrow} + d \sin \eta t$.
Note that $\eta \ll |\delta_{\rightarrow}|$, and at the 
same time $1/\eta$ is sufficiently smaller than $\tau_F$. 
This means that we can employ the separation of scales to 
take into account finite laser linewidth. We calculate the resulting 
force $F_y(\delta_\uparrow)$ on the cloud 
by averaging $F_y'(\delta_\rightarrow,\delta_\leftarrow,\delta_\uparrow)$ 
with the appropriate laser profile distribution in the frequency domain:
\begin{equation}
F_y\left(\delta_\uparrow\right)= \int_{-\infty}^{\infty} 
p(\omega; \omega_{01}+\delta_\rightarrow) 
F_y'(\omega-\omega_{01},
\omega-\omega_{01}+2|\delta_\rightarrow|,
\delta_\uparrow ) d\omega.
\end{equation}
The width $\sigma$ was determined to be $2\pi\times 1.5$~MHz 
by fitting to the experimental profiles. 
The Rabi frequencies used in the calculation are discussed 
in the Results Section.



\begin{thebibliography}{99}

\bibitem{Madison2000}
Madison, K.W., Chevy, F., Wohlleben, W., and Dalibard, J.,
Vortex Formation in a Stirred Bose-Einstein Condensate,
{\em Phys. Rev. Lett.} {\bf 84}, 806-809 (2000). 

\bibitem{Abo2001}
Abo-Shaeer, J.R., Raman, C., Vogels, J.M., and Ketterle, W., 
Observation of Vortex Lattices in Bose-Einstein Condensates, 
{\em Science} {\bf 292}, 476-479 (2001).

\bibitem{Lin2009}
Lin, Y-J., Compton, R.L., Jiménez-García, K., Porto, J.V., Spielman, I.B., 
Synthetic magnetic fields for ultracold neutral atoms, 
{\em Nature} {\bf 462}, 628-632 (2009). 

\bibitem{LeB2012} 
Le Blanc, L.J. {\em et al.}, 
Observation of a superfluid Hall effect, 
{\em PNAS } {\bf 109}, 10811-10814 (2012).

\bibitem{Aidelsburger2011}
Aidelsburger, M {\em et al.}, 
Experimental Realization of Strong Effective Magnetic Fields in an Optical Lattice,
{\em Phys. Rev. Lett.} {\bf 107}, 255301 (2011).

\bibitem{Struck2012}
Struck, J. {\em et al.}, 
Tunable Gauge Potential for Neutral and Spinless Particles in Driven Optical Lattices,
{\em Phys. Rev. Lett.} {\bf 108} 225304 (2012). 

\bibitem{Miyake2013}
Miyake, H., Siviloglou, G.A., Kennedy, C.J., Burton, W.C., and Ketterle, W., 
Realizing the Harper Hamiltonian with Laser-Assisted Tunneling in Optical Lattices, 
{\em Phys. Rev. Lett.} {\bf 111}, 185302 (2013).

\bibitem{Aidelsburger2013}
Aidelsburger, M. {\em et al.},
Realization of the Hofstadter Hamiltonian with Ultracold Atoms in Optical Lattices,
{\em Phys. Rev. Lett.} {\bf 111}, 185301 (2013).

\bibitem{Aidelsburger2014}
Aidelsburger, M. {\em et al.},
Measuring the Chern number of Hofstadter bands with ultracold bosonic atoms,
{\em Nature Physics} {\bf 11}, 162-166 (2015). 

\bibitem{Jotzu2014}
Jotzu, G. {\em et al.},
Experimental realization of the topological Haldane model with ultracold fermions, 
{\em Nature} {\bf 515}, 237-240 (2014). 

\bibitem{Ray2014}
Ray, M.W., Ruokokoski, E., Kandel, S., Mottonen, M., and Hall, D.S., 
Observation of Dirac monopoles in a synthetic magnetic field,
{\em Nature} {\bf 505}, 657-660 (2014). 


\bibitem{Bloch2012}
Bloch, I., Dalibard, J., and Nascimbene, S.,
Quantum simulations with ultracold quantum gases,
{\em Nature Physics} {\bf 8}, 267-276 (2012). 

\bibitem{Dal2011}
Dalibard, J., Gerbier, F., Juzeliunas, G., \"{O}hberg, P., 
Colloquium: Artificial gauge potentials for neutral atoms, 
{\em Rev. Mod. Phys.} {\bf 83}, 1523-1543 (2011). 

\bibitem{Goldman2014}
Goldman, N., Juzeliunas, G., \"{O}hberg, P., Spielman, I.B.,
Light-induced gauge fields for ultracold atoms, 
{\em Rep. Prog. Phys.} {\bf 77}, 126401 (2014). 

\bibitem{Cooper2008}
N.R. Cooper, 
Rapidly rotating atomic gases, 
{\em Adv. Phys.} {\bf 57}, 539-616 (2008). 


\bibitem{Dum1996}
Dum, R. and Olshanii, M.
Gauge Structures in Atom-Laser Interaction: Bloch Oscillations in a Dark Lattice, 
{\em Phys. Rev. Lett.} {\bf 76}, 1788-1891 (1996). 

\bibitem{Dubcek2014}
Dub\v{c}ek, T. {\em et al.},
Synthetic Lorentz force in classical atomic gases via Doppler effect and radiation pressure, 
{\em Phys. Rev. A} {\bf 89}, 063415 (2014). 

\bibitem{Met1999}
Metcalf, H.J., and Van Der Straten, P.,  
Laser Cooling and Trapping, 
(Springer, New York, 1999). 

\bibitem{Steck}
Steck, D.A., 
Rubidium 87 D Line Data. 
Available online at: 
http://steck.us/alkalidata/rubidium87numbers.pdf (2010)
Date of access: 1st June 2015.

\bibitem{Sheng2008}
Sheng, D., Perez Galvan, A., and Orozco, L.A.,  
Lifetime measurements of the 5d states of rubidium, 
{\em Phys. Rev. A} {\bf 78}, 062506 (2008).

\bibitem{Kaiser}
Chabe, J., {\em et al.}, 
Coherent and incoherent multiple scattering, 
{\em Phys. Rev. A} {\bf 89}, 043833 (2014).

\bibitem{Cheneau2008}
Cheneau, M., {\em et al.}, 
Geometric potentials in quantum optics: A semi-classical interpretation, 
{\em Europhys. Lett.} {\bf 83}, 60 001 (2008).



\end{thebibliography}

\section*{Acknowledgments}

This work was supported by the Unity through Knowledge Fund (UKF Grant No. 5/13).

\section*{Author contribution statement} 

All authors contributed in designing experiments, 
interpretation of the results, and writing the manuscript.
N.\v{S}. and T.B. performed all experiments. 
T.D., D.A., and H.B. performed all calculations. 



\end{document}